\documentclass[12pt]{article}
\usepackage{times}
\usepackage{geometry}
\usepackage[utf8]{inputenc}
\usepackage{enumitem,amssymb}
\usepackage{ragged2e,epstopdf}
\newlist{thematic}{itemize}{8}
\setlist[thematic]{label=$\square$}
\usepackage{pifont}

\usepackage{graphicx,amssymb,multirow,gensymb,lscape}
\usepackage{fancyhdr,aastex_hack,setspace,multicol}
\usepackage{wrapfig}
\usepackage{natbib}  
\usepackage{titlesec}

\titleformat*{\section}{\large\bfseries}
\titleformat*{\subsection}{\normalsize\bfseries}

\newcommand{\um}{\mbox{$\mu$m}}

\newcommand{\Msun}{\mbox{M$_{\odot}$}}

\begin{document}

\begin{center}
\Large\textbf{The Life Cycle of Dust} \\[0.5cm]
\large \textbf{Astro2020 Science White Paper}\\[1cm]
\normalsize
\end{center}

\noindent
\noindent \textbf{Thematic Areas:}  \hspace{5mm} $\square$ Planetary Systems \hspace{5mm} $\square$ Cosmology and Fundamental Physics \\
$\square$ Formation and Evolution of Compact Objects \hspace{5mm} $\boxtimes$ Star and Planet Formation  \\
$\boxtimes$ Stars and Stellar Evolution \hspace{5mm} $\square$ Resolved Stellar Populations and their Environments  \\
$\boxtimes$ Galaxy Evolution \hspace{5mm} $\square$ Multi-Messenger Astronomy and Astrophysics\\

\vspace{3mm}

\noindent \textbf{Principal authors:}\\[2mm]
Sarah I. Sadavoy\\
Harvard-Smithsonian Center for Astrophysics\\
sarah.sadavoy@cfa.harvard.edu\\
+1 617-496-7662\\

\noindent Mikako Matsuura\\
Cardiff University\\
MatsuuraM@cardiff.ac.uk\\
+44 29 2087 5120\\[2mm]

\noindent \textbf{Co-authors:}\\[2mm]
Lee Armus (Caltech/IPAC),
Cara Battersby (U. Connecticut), 
Caitlin Casey (U. Texas),
Christopher Clark (STScI),
Asantha Cooray (UC Irvine),
Karine Demyk (IRAP),
Neal Evans (U. Texas),
Karl Gordon (STScI),
Fr\'{e}d\'{e}ric Galliano (CEA),
Maryvonne Gerin (Observatoire de Paris), 
Benne Holwerda (U. Louisville), 
Nia Imara (CfA), 
Doug Johnstone (NRC), 
Alvaro Labiano (CSIC/INTA), 
David Leisawitz (NASA-Goddard),
Wanggi Lim (USRA), 
Leslie Looney (U. Illinois),
Margaret Meixner (STScI),
Eric Murphy (NRAO),
Roberta Paladini (Caltech),
Julia Roman-Duval (STScI),
Karin Sandstrom (UC San Deigo),
John-David Smith (U. Toledo),
Ian Stephens (CfA/SAO),
Nathalie Ysard (IAS)
\\[2mm]

\noindent \textbf{Abstract:}  Dust offers a unique probe of the interstellar medium (ISM) across multiple size, density, and temperature scales.  Dust is detected in outflows of evolved stars, star-forming molecular clouds, planet-forming disks, and even in galaxies at the dawn of the Universe.  These grains also have a profound effect on various astrophysical phenomena from thermal balance and extinction in galaxies to the building blocks for planets, and changes in dust grain properties will affect all of these phenomena.  A full understanding of dust in all of its forms and stages requires a multi-disciplinary investigation of the dust life cycle.  Such an investigation can be achieved with a statistical study of dust properties across stellar evolution, star and planet formation, and redshift.  Current and future instrumentation will enable this investigation through fast and sensitive observations in dust continuum, polarization, and spectroscopy from near-infrared to millimeter wavelengths.

\newpage


\section{Introduction} 

\vspace{-1mm}

\noindent Interstellar dust offers an unique opportunity to probe a wide range of astrophysical processes.  Even though dust makes up only a small portion of the interstellar medium (ISM) content (e.g., the Galactic dust-to-gas mass ratio is $\sim 1$\%), dust is present across multiple scales from diffuse ISM clouds to the ejecta of evolved stars and circumstellar disks \citep{Draine09}.  Thus, dust is a key tracer of galaxy emission, stellar evolution, and star and planet formation.\\[-2mm]

\begin{figure}[h!]
\includegraphics[width=0.98\textwidth,trim=0mm 0mm 0mm 0mm,clip=true]{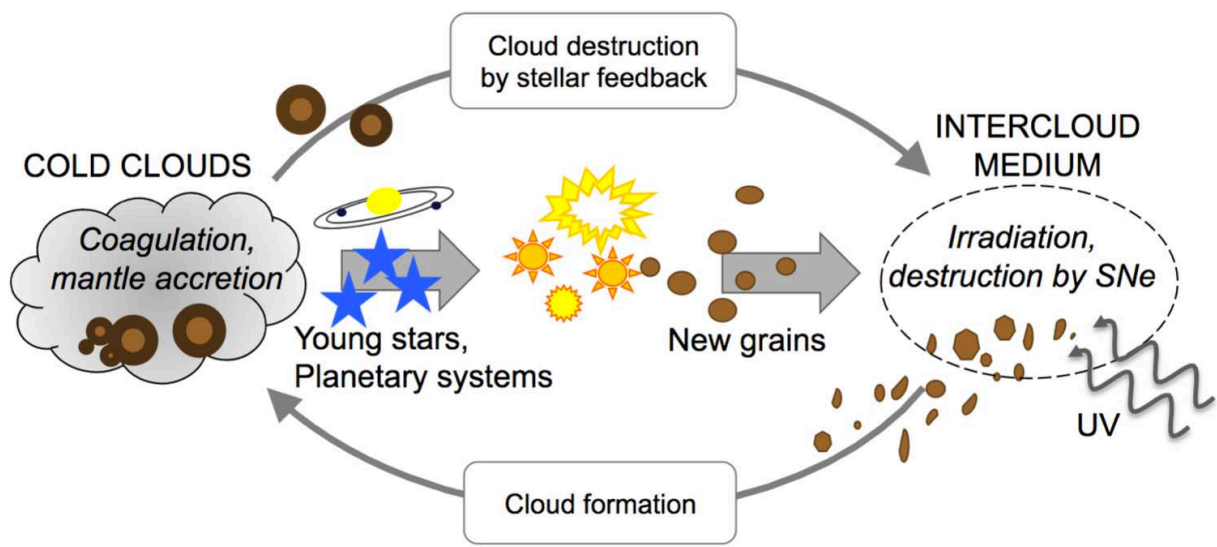}
\vspace{-2mm}
\caption{\emph{The life cycle of dust from \citep{ZhukovskaHenning14}. See text for details.}}\label{cartoon}
\end{figure}

\noindent Figure \ref{cartoon} outlines the life cycle of dust.  Briefly, dust refers to solid particles of silicates (Mg-rich or Fe-rich SiO$_n$) and carbonaceous materials with various degrees of hydrogenation and aromaticity, such as polycyclic aromatic hydrocarbons (PAHs).  These elements are produced in the interiors of stars and released into the ISM during the final stages of stellar evolution.  Dust in the ISM can be processed by shocks from supernova events or strong radiation.  Nevertheless, interstellar dust can still collect into clouds that are then cooled by their radiative processes.  The production of cold, dense clouds is paramount for star formation, and in these environments, dust grains will grow in size through coagulation and form the seeds for planets \citep[e.g.,][]{Draine03,Dwek09}.   The cycle begins again when the next generation of stars evolve off the main sequence.   \\[-5mm]

\noindent Dust grain sizes, compositions, and structures are expected to change throughout the life cycle, and these changes can have a profound effect on various astrophysical phenomena.   Grain properties influence (I) opacity and extinction, (II)  masses inferred from thermal dust emission, (III) how H$_2$ molecules form on grain surfaces, (IV) coupling efficiencies with an external magnetic field, (V) gas and ice chemistry on grain surfaces, and (VI) radiative coupling between UV photons and neutral gas \citep[e.g.,][]{Draine03, Andersson15}.  Astrophysical models need to reproduce all of these dependencies, which is a challenging task \citep[e.g.,][]{Zubko04, Ysard18}.  In particular, dust extinction observations are limited to nearby Galactic regions and often have coarse resolution based on the numbers of background stars, thermal dust emission maps are limited by sensitivity and uncertain dust properties (e.g., dust opacity), and scattering or polarization are limited to few, smaller regions.  These shortcomings limit our understanding of the dust life cycle.

\section{Main Science Themes}

\vspace{-3mm}

\subsection{The origins of interstellar dust}

\hspace{1mm}\\
\hspace{1mm}\\[-15mm]

\begin{wrapfigure}{r}{0.49\textwidth}
\centering
\vspace{-16mm}
\includegraphics[width=0.48\textwidth,trim=4mm 5mm 8mm 6mm,clip=true]{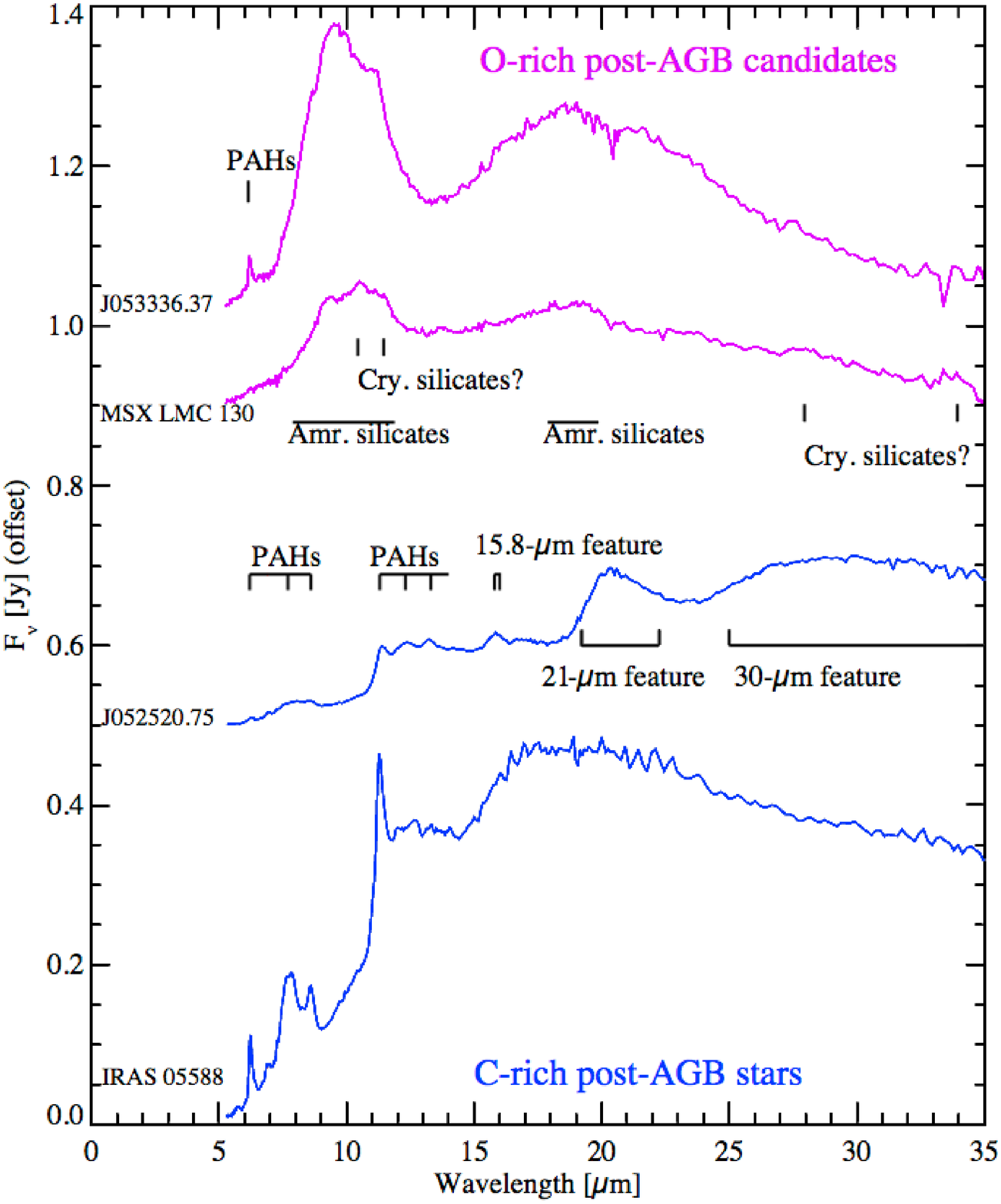}
\vspace{-4.5mm}
\caption{\emph{Spitzer dust spectra for oxygen-rich (purple) and carbon-rich (blue) AGB stars \citep[modified from][]{Matsuura14}.    } \label{evolved_stars}}
\end{wrapfigure}
\noindent How interstellar dust grains first form is greatly debated.  Dust can form in the stellar winds of evolved stars, in the ejecta of supernovae (SNe), and they can condense in situ in the ISM \citep{Draine09}, but which mechanism dominates is under investigation.  For example, isotopically anomalous silicates and the abundance of interstellar metals cannot be fully explained without dust accretion in the ISM.  Roughly 15\% of these silicates and most interstellar metals are expected to form directly within ISM and not from stardust \citep[e.g.,][]{Draine09, Zhukovska18}.  The dust associated with each of these production mechanisms will vary in size, composition, and structure.  Figure \ref{evolved_stars} shows example spectra of oxygen-rich and carbon-rich asymptotic giant branch (AGB) stars.  Oxygen-rich stars primarily produce silicates, whereas carbon-rich stars primarily produce carbon dust like PAHs and SiC \citep[e.g.,][]{Jones12}.    

\hspace{1mm}\\
\hspace{1mm}\\[-13mm]

\noindent  Interstellar dust has spectral features for both silicate and carbon grains.   Figure \ref{evolved_stars} illustrates several features, including silicate bands at $\sim$ 9.8 \um\ and 18 \um\ and several bands between 5-15 \um\ from PAHs.   The shape and position of these spectral bands reveal the dust structure and composition.  Broad, featureless lines indicate amorphous (disordered) molecular structure, whereas discrete narrow features indicate crystalline (ordered) structures \citep{Henning10}.  In the Milky Way, at least 95\%\ of interstellar silicate dust is amorphous and roughly 15\%\ of interstellar carbon is in PAHs \citep{Draine03}.  More active galaxies appear to have higher crystallinity \citep{Spoon06}.   \\[-3mm]

\noindent Dust production in a complex, multi-phase ISM is also closely linked to dust destruction.  Dust reformation in the ISM will produce primarily amorphous silicates or carbon-based materials that are able to condense and grow under constant UV photoexcitation \citep{Draine09}.  Crystalline silicates are likely produced from thermal processes in stellar sources.  For example, AGB winds show silicate crystallinity fractions of $\sim 10$\%\ that may correlate with outflow energy \citep{Jones12, Liu17}.  In the case of SNe, shocks can sputter and shatter interstellar dust and preferentially remove metals \citep[][white paper by M. Matsuura]{JonesNuth11, Matsuura19}, thereby shaping dust mineralogy and spectral line features \citep{deVries15}.     Thus, we must study how stardust and ISM dust each form and how long those grains will last in the ISM. \\[-8mm]

\subsection{Dust as a tracer of star and planet formation}

Stars and planets form in dense cores within molecular clouds, where grains are expected to  grow in size from micron sizes to planets in $\sim 10^{6-7}$ yr \citep[e.g.,][]{BlumWurm08, Testi14}.  Indeed, dust scattering at mid-infrared (MIR) wavelengths around cores and at (sub)millimeter wavelengths in protoplanetary disks indicate grains may reach sizes of up to $\sim 1$ \um\ and $\sim 100~\um$, respectively \citep[e.g.,][]{Pagani10,Kataoka15}.  Numerical models of dust coagulation in disks \citep[e.g.,][]{Birnstiel10} indicate that millimeter-sized dust can form quickly in the inner 10 au, but there are  few observational constraints. \\[-3mm]

\noindent Dust growth affects its opacity, a fundamental quantity that is necessary to convert thermal emission into mass.   Figure~3 shows how larger dust grains in cores have higher opacities than ISM dust \citep[e.g.,][]{Ossenkopf94, Ormel11, Kohler15}, and these opacities are expected to increase further from grain growth in disks \citep{DAlessio01, Testi14}.  Measuring accurate dust opacities is a significant challenge, however.  Dust opacity can vary with grain structure, metallicity, and ice content \citep[e.g.,][]{Demyk17, Ysard18}, and measurements are complicated by degeneracies with dust temperature \citep{Shetty09, Kelly12} or uncertainties from high optical depths, particularly in disks.  Due to these challenges, most studies assume a dust opacity value based on models, and these values can vary by factors of a few.  Improvements in dust opacity through direct observations are crucial to assess masses, gravitational stabilities and star formation potentials of clouds, cores, and disks.        \\[-9mm]

\begin{figure}[h!]
\begin{tabular}{lp{8.4cm}}
\includegraphics[width=0.33\textwidth,trim=0.5mm 0mm 0mm 0mm,clip=true,angle=-90]{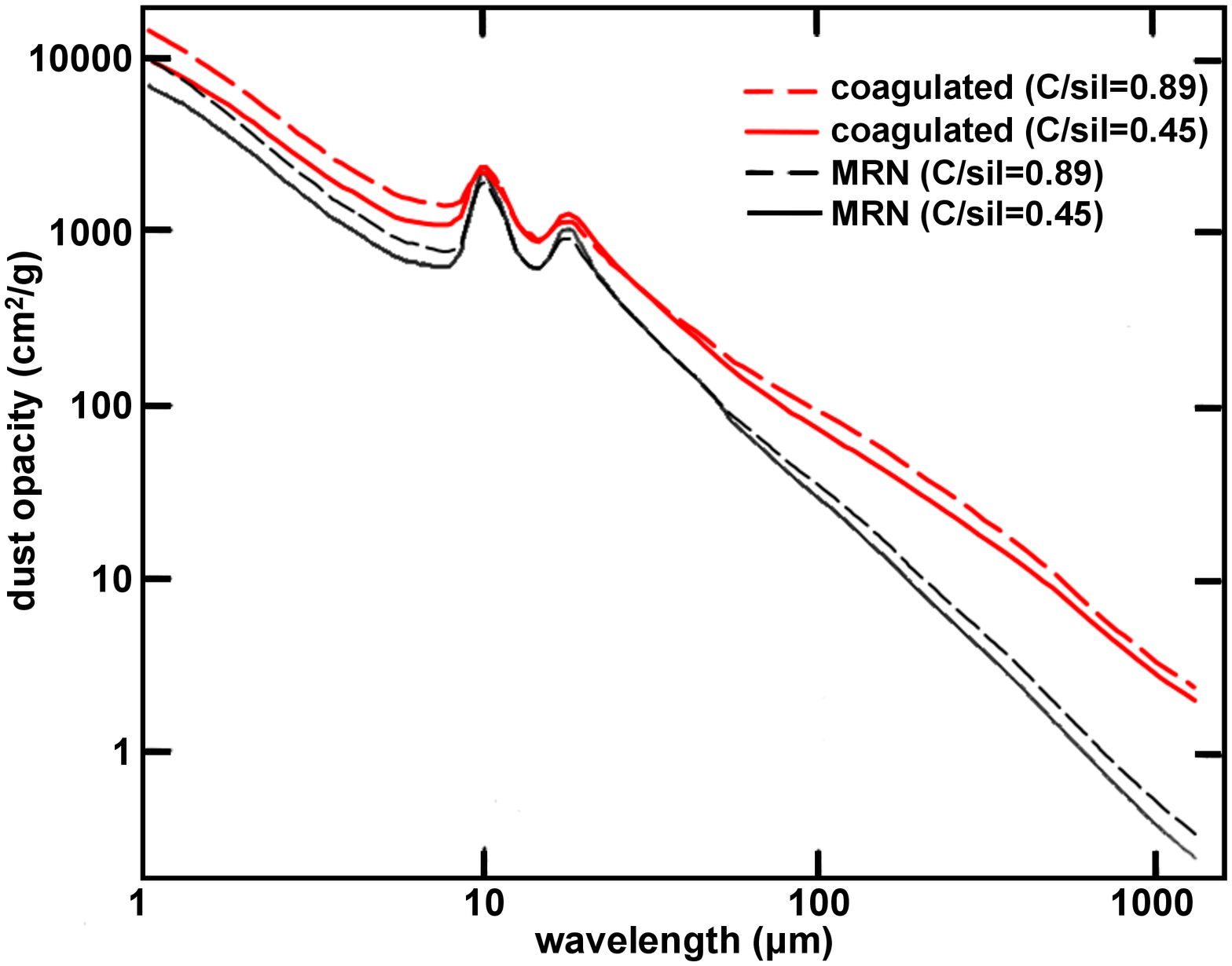} & \vspace{4mm} Figure 3: \emph{Models of dust opacity for gas densities typical of dense cores.  These curves compare ISM-like dust (MRN) with coagulated dust (after 10$\mathit{^5}$ yr at a gas density of 10$\mathit{^6}$ cm$\mathit{^{-3}}$), and at different ratios of carbonaceous dust (C) to silicate dust (sil).  In general, dust opacity is higher for larger dust grains and more carbon-rich dust.  This image was modified from \cite{Ossenkopf94}.  } \label{beta}\\[-4mm]
\end{tabular}
\end{figure}

\noindent Grain composition is also important in star and planet formation.  Crystallization fractions increase with evolutionary stage from protostars and disks \citep[e.g.,][]{Manoj11,Sturm13} to presolar grains in comets \citep[e.g.,][]{Roskosz15}.  Meteorites have crystalline fractions of $\sim 20$\%, which is substantially higher than the ISM fraction of $< 5$\% \citep{Draine09}.  Moreover, crystallinity may also depend on stellar mass \citep[e.g.,][]{Demyk99, Oliveira10} or location within a disk \citep{Watson09}.  Grain processing in disks will affect the formation of ice mantles, which is important for gas-grain reactions that produce complex (organic) molecules \citep[e.g.,][]{BerginTafalla07}.  The interplay between ices, grain growth, dust composition, and structure is  an important open question for the origin of life.   \\[-8mm]

\renewcommand\thefigure{\arabic{figure}}\setcounter{figure}{3}

\subsection{Dust grain properties across cosmic time}

Dust grains in high redshift ($z > 5$) galaxies may be vastly different from the dust in the Milky Way as these systems have had less time  ($\lesssim 1$ Gyr) to enrich their ISMs and allow for grain regrowth.  Nevertheless, some high-redshift galaxies have substantial dust emission \citep[e.g.,][]{Vieira13, Scoville16} and inferred dust reservoirs that are larger than the Milky Way \citep[$> 10^8$ \Msun; e.g.,][]{Marrone18}.  Although these inferred dust masses are unreliable due to incomplete knowledge of high-redshift dust \citep{Imara18}, the presence of large quantities of dust in galaxies at $z > 5$ indicates that star formation and dust production must occur very quickly.      

\noindent   Dusty, high redshift galaxies emit most of their energies at FIR wavelengths.   The origin and growth of high redshift dust is an open question, but must occur on relatively short timescales \citep{Draine09}.  Figure 4 shows a model for how early galaxies could obtain their dust masses.  There is significant debate about the contribution of AGB stars at high redshifts, with some studies saying that core-collapse SNe should dominate \citep[e.g.,][]{Galliano08, Nozawa09} with others finding that AGB stars could be significiant if early galaxies have high star formation rates ($> 1000$ \Msun\ yr$^{-1}$) and top-heavy initial mass functions \citep[][]{Valiante09, DwekCherchneff11}.  Dust from AGB is important for the production of carbon dust (e.g., PAHs), as these grains are necessary for UV absorption in the spectra of high redshift galaxies \citep{Casey14}.       \\[-9mm]

\begin{figure}[h!]
\begin{tabular}{lp{7.5cm}}
\includegraphics[width=0.38\textwidth,angle=-90,trim=0mm 0mm 0mm 2mm,clip=true]{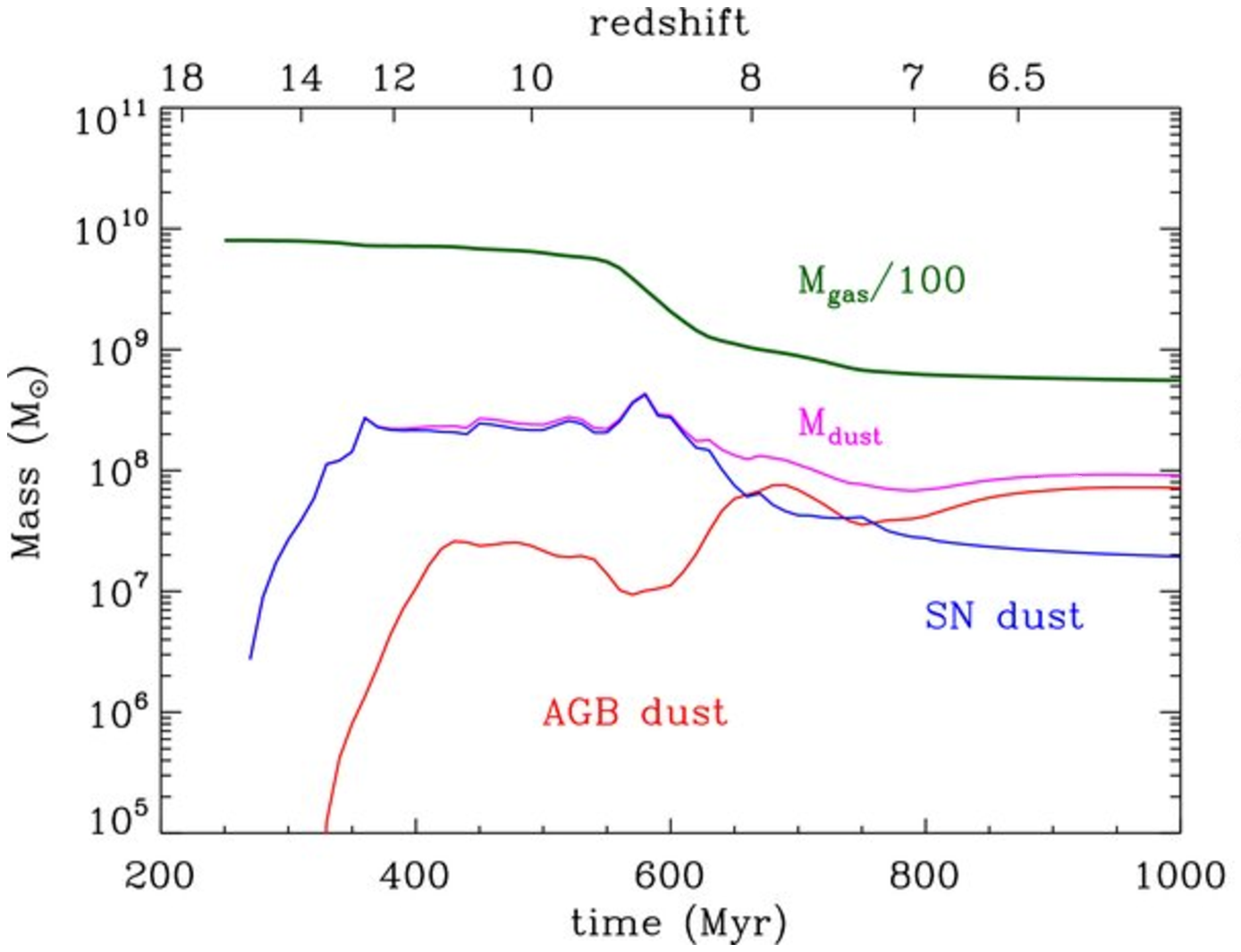} & \vspace{3mm} Figure 4: {\emph{Evolution of dust mass at early epochs for a starburst galaxy from \citet{DwekCherchneff11}.  The curves show the total dust mass (purple) and the relative contributions from AGB dust (red) and SN dust (blue).  The corresponding gas mass in the model is shown in green.   Note that several exceptionally dusty high-redshift ($z~>~6$) galaxies have inferred dust masses of $\mathit{10^{8-9}}$~\Msun\ \citep{Bertoldi03, Beelen06, Marrone18}.}}\\[-4mm]
\end{tabular}
\end{figure}

\noindent Questions about the composition of dust at high redshifts can be addressed by comparing those systems with more nearby galaixes, which exhibit a range of properties from low-metallicities to starbursts \citep[e.g.,][]{Matsuura07, Galliano18}.  Moreover, observations of a variety of environments within the Milky Way (e.g., Galactic center, high latitude clouds) offer the chance to spatially resolve dusty structures and determine the conditions surrounding the formation of the first stars. At present, such observations of local dust are piecemeal or incomplete.  For example, there are limited observational studies of dust produced by SN in low-metallicity environments or of dust composition at high redshifts.  Key spectral features are shifted to MIR and FIR wavelengths and require highly sensitive detectors to identify spectral signatures.   \\[-8mm]

\section{Future Prospects}

\vspace{-2mm}

\noindent A proper investigation of the life cycle of dust must combine laboratory measurements and models with observations.  Laboratories provide the necessary templates to identify dust compositions and can trace how grain properties under different conditions affect dust opacities  \citep[e.g.,][]{Koike10, Demyk17}.  Models test dust extinction, polarization, and scattering processes \citep[e.g.,][]{Ysard18}.  Large, unbiased, statistical studies of dust throughout the ISM in different environments within the Milky Way and out to high redshifts are still needed to fully understand the dust life cycle.  Here we focus on observations of dust composition and dust sizes. \\[-2mm]  

\noindent \textbf{Dust composition and structure:} Investigations of dust composition and structure across the life cycle require large, systematic spectroscopic studies of dust composition in evolved stars and stellar remnants, molecular clouds, protostellar disks, and the ISM of nearby and more distant galaxies.  Comparisons between dust composition in different environments locally (e.g., low metallicity, high density and pressure) are especially important to study how dust is produced and how those grains are processed at high redshift.  The formation of dust in early universe is critical to H$_2$ formation in metal-poor clouds and thermal balance at early times when there are few metals.    \\[-2mm]

\noindent Key silicate, carbonaceous, and ice features are found at wavelengths of a few micron to $\sim 90$~\um\ \citep[e.g.,][]{Henning10}.  Previous observations at these wavelengths were possible with \emph{Spitzer} (primarily at $\sim 3-30$ \um) and \emph{Herschel} (at $\gtrsim 60$ \um), and current observations are possible with SOFIA (at $\gtrsim 5$ \um).  But studies with these facilities are often piecemeal and mainly target bright, nearby sources \citep[e.g.,][]{Kemper02, Oliveira10, Matsuura19} that may not reflect a range of different environment conditions.  Future observations from the \emph{JWST} will have the sensitivity to enable dust composition studies in more extreme environments at $z \lesssim 3$ and toward the centers of starburst galaxies, and proposed missions like the \emph{OST} and \emph{SPICA} will offer high sensitivity studies of dust composition in the Milky Way and $z \gtrsim 1$. \\[-2mm]

\noindent \textbf{Dust sizes:}  Grain sizes inferred from dust opacity curves require multi-wavelength FIR observations.  While there are several large FIR Galactic and extragalactic surveys with \emph{Herschel} \citep[e.g.,][]{Andre10, Maddox18}, these studies were limited to a maximum wavelength of 500 \um\ and cannot break the degeneracy with dust temperature.  Several approaches have been developed to combine \emph{Herschel} observations with longer wavelength data to circumvent the degeneracy \citep{Sadavoy13, Forbrich15,Sadavoy16}, but these studies have been limited to the small regions of nearby Galactic clouds.  Observations from \emph{Planck} can probe dust opacity curves across the entire sky at $\gtrsim 500$ \um\ \citep{Planck_beta15}, but at coarse ($> 5\arcmin$) resolutions; grain growth primarily occurs on much smaller scales. \\[-2mm]

\noindent New detectors such as NIKA-2 and TolTEC have fast mapping ($\sim 10$ deg$^2$ mJy$^{-2}$ h$^{-1}$) capabilities at $\lambda > 1$ mm at high resolutions ($\lesssim 10$\arcsec) to probe dust opacities and grain growth in distant Galactic clouds and other galaxies (see white paper by C. Clark).  Dust opacities will be further improved with MIR and FIR data (e.g., from \emph{Spitzer}, \emph{Herschel}, or proposed missions like \emph{OST} and \emph{SPICA}) to model cold and warm dust \citep[e.g.,][]{MeisnerFinkbeiner15} and dust composition.   Multi-wavelength dust polarization capabilities are further needed to study dust alignment efficiencies.  Larger dust grains and iron-poor dust are less efficient at aligning with magnetic fields via radiative torques and will have weaker polarization \citep{Andersson15, Tazaki17}.  In disks, grain sizes can be constrained via polarized dust scattering \citep{Kataoka15}.  ALMA has revolutionized this approach, although multi-wavelength observations are needed to disentangle different mechanisms \citep[][white paper by I. Stephens]{Stephens17, Yang17}. ALMA can also probe how grain growth affects gas chemistry, although other facilities (e.g., \emph{JWST}, \emph{OST}) are needed to study dust and ice features which are primarily at MIR wavelengths.  \\[-2mm]     
 
 \noindent  \textbf{Conclusions:} While there are many studies about dust in galaxies and the Milky Way, we lack facilities that can investigate the entire dust life cycle with large, uniform, and statistically significant surveys.  Specifically, we require that the next generation of detectors have wide-field continuum and spectroscopic capabilities from MIR to FIR wavelengths.  With such observations, we can uncover dust properties in the Milky Way, nearby galaxies, and the early Universe, and investigate not only how dust grains are produced, but also the timescales for them to grow and evolve in different environments.  These observations are paramount to produce accurate dust models and make predictions for many astrophysical processes, ranging from planet formation to galaxy evolution.   
 
\pagebreak

\newpage

\twocolumn

\setlength{\bibsep}{5pt}


\end{document}